\documentclass[10pt,conference]{IEEEtran}
\IEEEoverridecommandlockouts\IEEEpubid{\makebox[\columnwidth]{ J. Xu is the corresponding author. \hfill} \hspace{\columnsep}\makebox[\columnwidth]{ }}

\ifCLASSOPTIONcompsoc
\usepackage[caption=false,font=normalsize,labelfon
t=sf,textfont=sf]{subfig}
\else
\usepackage[caption=false,font=footnotesize]{subfi
g}
\fi

\usepackage{amsmath,amssymb}
\usepackage{multicol}
\usepackage{graphicx,graphics,color,psfrag}
\usepackage{cite,balance}
\usepackage{algorithm}
\usepackage{accents}

\usepackage{bm}
\usepackage{url}
\usepackage{algorithmic}
\usepackage[english]{babel}
\usepackage{multirow}
\usepackage{enumerate}
\usepackage{cases}
\usepackage{stfloats}
\usepackage{dsfont}
\usepackage{color,soul}
\usepackage{amsfonts}
\usepackage{tcolorbox}
\usepackage{amsmath}
\usepackage{float}

\usepackage{cite,graphicx,amsmath,amssymb}
\usepackage{fancyhdr}
\usepackage{hhline}
\usepackage{graphicx,graphics}
\usepackage{array,color}
\usepackage{amsmath}
\usepackage{stfloats}
\usepackage[flushleft]{threeparttable}
\usepackage{booktabs}

\newtheorem{proposition}{Proposition}
\newtheorem{remark}{Remark}

\ifCLASSINFOpdf

\else

\fi

\begin{document}
\title{Cram\'er-Rao Bound Minimization for IRS-Enabled Multiuser Integrated Sensing and Communication with Extended Target}

\author{
\IEEEauthorblockN{Xianxin~Song\IEEEauthorrefmark{1}, Tony~Xiao~Han\IEEEauthorrefmark{2}, and  Jie~Xu\IEEEauthorrefmark{1}%
}
\IEEEauthorblockA{\IEEEauthorrefmark{1}School of Science and Engineering (SSE) and Future Network of Intelligence Institute (FNii),\\ The Chinese University of Hong Kong (Shenzhen), Shenzhen,  China}
\IEEEauthorblockA{\IEEEauthorrefmark{2}Wireless Technology Lab, 2012 Laboratories, Huawei, Shenzhen, China}
Email: xianxinsong@link.cuhk.edu.cn, tony.hanxiao@huawei.com, xujie@cuhk.edu.cn% <-this % stops a space
}

\maketitle
\begin{abstract}
This paper investigates an intelligent reflecting surface (IRS) enabled multiuser integrated sensing and communication (ISAC) system, which consists of one multi-antenna base station (BS), one IRS, multiple single-antenna communication users (CUs), and one extended target at the non-line-of-sight (NLoS) region of the BS. The IRS is deployed to not only assist the communication from the BS to the CUs, but also enable the BS's NLoS target sensing  based on the echo signals from the BS-IRS-target-IRS-BS link. To provide full degrees of freedom for sensing, we suppose that the BS sends additional dedicated sensing signals combined with the information signals. Accordingly, we consider two types of CU receivers, namely Type-I and Type-II receivers, which do not have and have the capability of cancelling the interference from the sensing signals, respectively. Under this setup, we jointly optimize the transmit beamforming at the BS and the reflective beamforming at the IRS to minimize the Cram\'er-Rao bound (CRB) for estimating the target response matrix with respect to the IRS, subject to the minimum signal-to-interference-plus-noise ratio (SINR) constraints at the CUs and the maximum transmit power constraint at the BS. We present efficient algorithms to solve the highly non-convex SINR-constrained CRB minimization problems, by using the techniques of alternating optimization and semi-definite relaxation. Numerical results show that the proposed design achieves lower estimation CRB than other benchmark schemes, and the sensing signal interference pre-cancellation is beneficial when the number of CUs is greater than one.
\end{abstract}

\IEEEpeerreviewmaketitle

\section{Introduction}
Integrated sensing and communication (ISAC) has been recognized as one of the key technologies for sixth-generation (6G) wireless networks to provide ubiquitous sensing and communication services by reusing base station (BS) infrastructures and spectrum resources (see, e.g., \cite{liu2021integrated,9705498} and the references therein). In practice, the performance of ISAC networks is highly dependent on the wireless propagation environment. Due to the obstacles such as buildings in outdoor scenarios and furniture/walls in indoor scenarios, the direct links between BSs and sensing targets/communication users (CUs) may be weakened or even blocked, thus seriously degrading the sensing and communication performances.

Recently, intelligent reflecting surface (IRS) has emerged as a promising technology to resolve the above issues by reconfiguring the wireless environment via properly adjusting the phase shifts of its passive reflecting elements. As such, IRSs can improve the communication performance by, e.g., enhancing the received signal strength and refining channel ranks \cite{8811733,9122596,9326394}, and can also enhance the wireless sensing performance via creating new virtual line-of-sight (LoS) links and providing additional sensing angles \cite{Stefano,xianxin}. 

Despite the benefits, IRS-enabled ISAC introduces new technical issues to be dealt with. First, as compared to the counterpart without IRSs, the reflective beamforming at IRSs (via controlling the reflection coefficients) becomes a new design degree of freedom (DoF) for enhancing the system performance. This, however, is particularly challenging, due to various practical considerations such as the unit-modulus reflection constraints \cite{8811733}. Next, as sensing and communication tasks coexist in the same system, how to optimally balance their performance tradeoff via the joint transmit and reflective beamforming design is another important task. This, however, is difficult, due to their complicated coupling relationship. 

In the literature, there have been several prior works investigating IRS-enabled ISAC under different setups. For instance, the authors in \cite{song2021joint} considered an IRS-enabled ISAC system with one BS, one IRS, one CU, and multiple point targets at the non-LoS (NLoS) region of the BS, in which the BS uses the virtual LoS links (i.e., the BS-IRS-targets-IRS-BS links) for sensing, and the joint beamforming was optimized to maximize the minimum sensing beampattern gain (or worst-case target illumination power) at targeted angles with respect to (w.r.t.) the IRS while ensuring the signal-to-noise ratio (SNR) requirement at the CU. The authors in \cite{sankar2022beamforming} then studied a more generic IRS-enabled ISAC system  with multiple CUs by considering two different IRS configurations. Furthermore, the authors in \cite{9769997} considered the IRS-enabled ISAC scenario with both the direct BS-target-BS and the reflected BS-IRS-target-IRS-BS links utilized for target sensing, in which the joint beamforming was optimized to maximize the radar signal-to-interference-plus-noise ratio (SINR), subject to various quality-of-service (QoS) requirements for communication. Besides using sensing beampattern gains and radar SINR as sensing performance measures, another line of IRS-enabled ISAC research \cite{9591331,wang2022stars} adopted the Cram\'er-Rao bound (CRB) for target estimation/tracking as the sensing performance metric. CRB provides a lower bound on the variance of any unbiased parameter estimators, and thus serves as the fundamental limits for the sensing performance \cite{bekkerman2006target,9652071}. 
However, prior work \cite{9591331} used IRS to assist communication only and performed sensing based on the direct BS-target-BS links, and \cite{wang2022stars} performed sensing at IRS based on the BS-IRS-target-IRS links via deploying additional receivers at the IRS. These works are not applicable to the scenario when the IRS is purely passive (without dedicated receivers) and the NLoS target sensing needs to be performed at the BS through the BS-IRS-target-IRS-BS link. This thus motivates our work in this paper. 

This paper considers an IRS-enabled multiuser ISAC system with one BS, one IRS, multiple CUs, and an extended target at the NLoS region of the BS. The IRS is deployed to not only assist the wireless communication from the BS to the CUs, but also create a virtual LoS link (i.e., the BS-IRS-target-IRS-BS link) to facilitate the target sensing at the BS. In order to achieve full DoFs for target sensing, we suppose that the BS transmits additional dedicated sensing signals combined with the information signals \cite{9124713,9652071,hua}. The dedicated sensing signals can be pseudorandom or deterministic sequences, which are known by the CUs prior to the transmission. As a result, we consider two types of CU receivers, namely Type-I and Type-II receivers, which do not have and have the capability of canceling the interference from the sensing signals, respectively. Furthermore, with an extended target, the BS aims to estimate the complete target response matrix w.r.t. the IRS based on the echo signals from the BS-IRS-target-IRS-BS link, for which the estimation CRB has been derived in \cite{xianxin}. Building upon this, we jointly optimize the transmit beamforming at the BS and the reflective beamforming at the IRS to minimize the estimation CRB, subject to the minimum SINR constraints at the CUs and the maximum transmit power constraint at the BS. Although the formulated SINR-constrained CRB minimization problems are highly non-convex, we present efficient algorithms to obtain converged solutions by using alternating optimization and semi-definite relaxation (SDR). Numerical results show that the proposed design achieves improved sensing performance in terms of lower CRB, as compared to other benchmark schemes. It is also shown that the case with Type-II CU receivers leads to significant performance gains over that with Type-I CU receivers when the number of CUs is greater than one, thus validating the benefit of sensing signal interference pre-cancellation.

\textit{Notations:} 
Boldface letters refer to vectors (lower case) or matrices (upper case). For a square matrix $\mathbf S$, $\mathbf S^{-1}$ denotes its inverse, and $\mathbf S \succeq \mathbf{0}$ means that $\mathbf S$ is positive semi-definite. For an arbitrary-size matrix $\mathbf M$, $\mathrm {rank}(\mathbf M)$, $\mathbf M^{T}$, and $\mathbf M^{H}$ denote its rank, transpose, and conjugate transpose, respectively. We use $\mathcal{C N}(\mathbf{x}, \mathbf{\Sigma})$ to denote the distribution of a circularly symmetric complex Gaussian (CSCG) random vector with mean vector $\mathbf x$ and covariance matrix $\mathbf \Sigma$, and $\sim$ to denote “distributed as”. The space of $x \times y$ complex matrices is denoted by $\mathbb{C}^{x \times y}$. The symbol $\|\cdot\|$ stands for the Euclidean norm, $|\cdot|$ for the magnitude of a complex number, and $\mathrm {diag}(a_1,\cdots,a_N)$ for a diagonal matrix with diagonal elements $a_1,\cdots,a_N$.
\begin{figure}[t]
    \centering
    \includegraphics[width=0.33\textwidth]{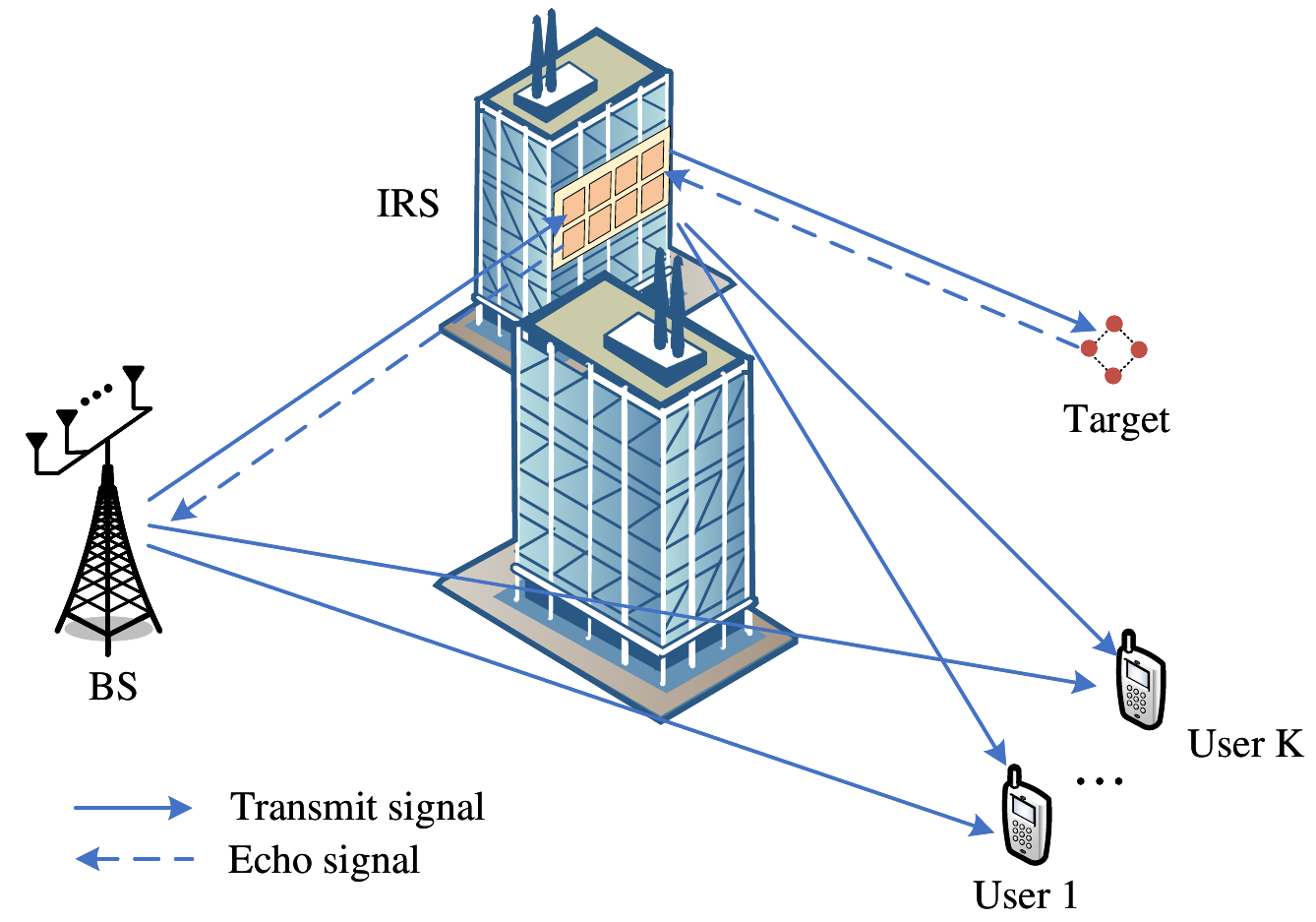}
    \caption{System model of IRS-enabled multiuser ISAC.}
    \label{system_model}
\end{figure}
\section{System Model}

We consider an IRS-enabled ISAC system as shown in Fig.~\ref{system_model}, which consists of one BS with $M>1$ antennas, one IRS with $N>1$ reflective elements, $K \ge 1$ single-antenna CUs, and an extended target at the NLoS region of the BS. The set of CUs and that of reflection elements at the IRS are denoted by $\mathcal{K}=\{1,\cdots,K\}$ and $\mathcal{N}=\{1,\cdots,N\}$, respectively.

We consider one particular ISAC transmission block consisting of $T$ symbols with $T$ being sufficiently large. Let $\mathcal{T} = \{1, \cdots, T\}$ denote the set of symbols. We suppose that the BS sends both information signals and dedicated sensing signals, for achieving full DoFs for sensing \cite{9124713,9652071,hua}. Let $s_k(t)$ denote the transmit information signal for the $k$-th CU at symbol $t$, and $\mathbf w_k$ denote the corresponding transmit beamforming vector. Here, $s_k(t)$'s are assumed to be independent and identically distributed (i.i.d.) random variables with zero mean and unit variance. Let $\mathbf x_0(t) \in\mathbb{C}^{M \times 1}$ denote the dedicated sensing signal vector at symbol $t$, which is independent of the information signals and its sample covariance matrix is $\mathbf R_0 \triangleq \frac{1}{T}\sum_{t\in\mathcal{T}}\mathbf{x}_0(t)\mathbf{x}_0^{H}(t) \succeq \mathbf{0}$. Then the transmitted signal by the BS at symbol $t$ is
\begin{equation}
\mathbf x(t) =\sum_{k\in\mathcal{K}}\mathbf w_k s_k(t) + \mathbf{x}_0(t), \forall t \in \mathcal{T}.
\end{equation} 
The sample (or statistical) covariance matrix of the transmitted signal is
\begin{equation}\label{eq:approx_R}\nonumber
\mathbf R_x \! \triangleq \! \frac{1}{T}\hspace{-0.2em}\sum_{t\in\mathcal{T}}\mathbf{x}(t)\mathbf{x}^{H}(t)
 \approx \mathbb{E}(\mathbf{x}(t)\mathbf{x}^{H}(t))= \hspace{-0.3em} \sum_{k\in\mathcal{K}} \mathbf w_k\mathbf w_k^{H} + \mathbf R_0,
\end{equation}
where the approximation holds as $T$ is sufficiently large. Let $P_0$ denote the maximum transmit power at the BS, i.e.,
\begin{equation}\label{equ:sum_power_constr}
  \mathbb{E}(\| \mathbf x(t) \|^2) =\sum_{k\in\mathcal{K}} \|\mathbf w_k\|^2 + \mathrm {tr}(\mathbf R_0) \le P_0.
\end{equation}
Furthermore, we consider that the IRS employs the reflective beamforming to facilitate the ISAC operation. In particular, the IRS can only adjust the phase shifts at the reflecting elements with unit amplitudes\cite{8811733}. Let $\mathbf v = [e^{j\phi_1},\ldots,e^{j\phi_{N}}]^{T}$ denote the reflective beamforming vector at the IRS, with $\phi_n \in (0, 2\pi]$ being the phase shift of each element.

First, we consider the wireless communication from the BS to the CUs. Let $\mathbf h_{\text{d},k}^{H}\in \mathbb C^{1\times M}$ and $\mathbf h_{\text{r},k}^{H}\in \mathbb C^{1\times N}$ denote the channel vectors from the BS and the IRS to the $k$-th CU, respectively. Let $\mathbf G \in \mathbb C^{N\times M}$ denote the channel matrix from the BS to the IRS. We assume the BS perfectly knows the channel state information (CSI) of all channels via proper channel estimation methods\cite{9722893}. The received signal by the $k$-th CU at symbol $t$ is
\begin{equation}\nonumber
\begin{split}
&y_k(t) =  (\mathbf h_{\text{d},k}^{H}+ \mathbf h_{\text{r},k}^{H} \mathbf \Phi\mathbf G )\mathbf x(t)+  n_k(t)\\
 &= \underbrace{(\mathbf h_{\text{d},k}^{H}+ \mathbf h_{\text{r},k}^{H} \mathbf \Phi\mathbf G )\mathbf w_k s_k(t)}_\text{desired information signal}\! + \underbrace{(\mathbf h_{\text{d},k}^{H}+ \mathbf h_{\text{r},k}^{H} \mathbf \Phi\mathbf G )\hspace{-0.6em}\sum_{{i\in\mathcal{K},i\neq k}}\hspace{-0.6em}\mathbf w_i s_i(t)}_\text{inter-user interference} \\
 &\quad +  \underbrace{(\mathbf h_{\text{d},k}^{H}+ \mathbf h_{\text{r},k}^{H} \mathbf \Phi\mathbf G )\mathbf x_0(t)}_\text{sensing signal interference} + n_k(t), \forall t \in \mathcal{T},  
 \end{split}
\end{equation}
where $n_k(t) \sim \mathcal{CN}(0,\sigma_k^2)$ is the additive white Gaussian noise (AWGN) at the $k$-th CU receiver, and $\mathbf{\Phi} = \mathrm {diag}(\mathbf v)$ is the reflection matrix of the IRS.
Note that the dedicated sensing signals $\mathbf x_0(t)$'s can be pseudorandom or deterministic sequences, that are generated offline  and known by the CUs prior to the transmission.
As a result, we consider two types of CU receivers, namely Type-I (legacy) and Type-II (dedicatedly designed) receivers, which can not pre-cancel and can pre-cancel the interference caused by the sensing signals, respectively. For the case with Type-I or Type-II CU receivers, the corresponding SINR at the $k$-th CU is 
\begin{equation}\label{eq:SINR}
\gamma_k^{(\text{I})}=\frac{|\mathbf h_k^{H} \mathbf w_k|^2}{\sum_{i\in\mathcal{K},i\neq k}|\mathbf h_k^{H}\mathbf w_i|^2+\mathbf h_k^{H}\mathbf R_0 \mathbf h_k +\sigma_k^2}, \forall k \in \mathcal{K},
\end{equation}
\begin{equation}\label{eq:SINR_II}
\gamma_k^{(\text{II})}= \frac{|\mathbf h_k^{H}\mathbf w_k|^2}{\sum_{i\in\mathcal{K},i\neq k}|\mathbf h_k^{H}\mathbf w_i|^2+ \sigma_k^2}, \forall k \in \mathcal{K},
\end{equation}
where $\mathbf h_k = \mathbf h_{\text{d},k}+ \mathbf G^{H} \mathbf \Phi^{H} \mathbf h_{\text{r},k}$ denotes the combined channel vector from the BS to the $k$-th CU. 

Next, we consider the sensing of an extended target, which consisting of multiple point-like scatterers in an extended region of space\cite{9652071,xianxin}. We denote  $\mathbf H$ as the target response matrix w.r.t. the IRS (i.e., the cascaded channel over the IRS-target-IRS link). In this case, the received echo signal vector at the BS at symbol $t$ is 
\begin{equation}\label{eq:echo_signal}
\mathbf y_\text{R}(t) = \mathbf G^{T} \mathbf \Phi^{T} \mathbf H \mathbf \Phi \mathbf G \mathbf x(t) + \mathbf n_\text{R}(t), \forall t \in \mathcal{T},
\end{equation}
where $\mathbf n_\text{R}(t)\sim \mathcal{C N}(\mathbf{0}, \sigma_\text{R}^2\mathbf I_M)$ denotes the AWGN at the BS receiver, which includes the noise at the BS and the clutter interference from environment. Based on \eqref{eq:echo_signal}, the sensing objective is to estimate the target response matrix $\mathbf H$ from $\mathbf y_\text{R}(t)$, in which $\mathbf x(t)$ (including both the information signals and the dedicated sensing signals), $\mathbf G$, and $\mathbf \Phi$ are known by the BS. 
As such, we use the CRB for estimating $\mathbf H$ as the sensing performance measure. Based on the derivation in \cite{xianxin}, we have the estimation CRB as  
\begin{equation}\label{eq:CRB_extended_R_x}
\text{CRB}(\mathbf H)=\frac{\sigma_\text{R}^2}{T}\mathrm{tr}((\mathbf G\mathbf R_x\mathbf G^{H})^{-1}) \mathrm{tr}((\mathbf G\mathbf G^{H})^{-1}).
\end{equation}

Our objective is to minimize the estimation CRB for sensing, i.e., $\text{CRB}(\mathbf H)$ in \eqref{eq:CRB_extended_R_x}, by jointly optimizing the transmit beamformers $\{\mathbf w_k\}$ and $ \mathbf R_0$ at the BS and the reflective beamformer $\mathbf v$ at the IRS, subject to the minimum SINR constraints at the CUs and the maximum transmit power constraint at the BS. Based on \eqref{eq:CRB_extended_R_x}, minimizing $\text{CRB}(\mathbf H)$ is equivalent to minimizing $\mathrm{tr}((\mathbf G\mathbf R_x\mathbf G^{H})^{-1})= \mathrm{tr}\big(\big(\mathbf G\big(\sum_{k\in \mathcal{K}}\mathbf w_k \mathbf w_k^H + \mathbf R_0\big)\mathbf G^H\big)^{-1}\big)$. As a result, the SINR-constrained CRB minimization problems with Type-I and Type-II CU receivers are formulated as (P1) and (P2), respectively. 
\begin{subequations}
\begin{align} \nonumber
  &\text{(P1)}:     \min_{\{\mathbf w_k\}, \mathbf R_0, \mathbf v}     \mathrm{tr}\left(\left(\mathbf G\left(\sum_{k\in\mathcal{K}} \mathbf w_k\mathbf w_k^{H} + \mathbf R_0\right)\mathbf G^{H}\right)^{-1}\right)  \\  \label{eq:SINR_minimum_I}
   &\text{s.t.} \  \frac{|\mathbf h_k^{H} \mathbf w_k|^2}{\sum_{i\in\mathcal{K},i\neq k}|\mathbf h_k^{H}\mathbf w_i|^2+\mathbf h_k^{H}\mathbf R_0 \mathbf h_k \!+\!\sigma_k^2}\ge \Gamma_k, \forall k\in \!\mathcal{K}\\ \label{eq:power}
  &\quad ~\sum_{k\in\mathcal{K}}\|\mathbf w_k\|^2  + \mathrm {tr}(\mathbf R_0) \le P_0 \\\label{eq:semi}
   &\quad ~\mathbf R_0 \succeq \mathbf{0}\\\label{eq:phase_1}
  &\quad ~\mathbf |\mathbf v_n|=1, \forall n\in \mathcal{N}.
  \\\nonumber
  \\
  &\text{(P2)}:    \nonumber \min_{\{\mathbf w_k\}, \mathbf R_0, \mathbf v}     \mathrm{tr}\left(\left(\mathbf G\left(\sum_{k\in\mathcal{K}} \mathbf w_k\mathbf w_k^{H} + \mathbf R_0\right)\mathbf G^{H}\right)^{-1}\right)    \\   \label{eq:SINR_minimum_II}
  ~ &\text{s.t.}  \ \   \frac{|\mathbf h_k^{H}\mathbf w_k|^2}{\sum_{i\in\mathcal{K},i\neq k}|\mathbf h_k^{H}\mathbf w_i|^2+ \sigma_k^2}\ge \Gamma_k, \forall k\in \mathcal{K} \tag{8}\\ 
  & \nonumber  \qquad \eqref{eq:power}, ~\eqref{eq:semi},~\text{and}~\eqref{eq:phase_1}.
\end{align}
\end{subequations}
In (P1) and (P2), $\Gamma_k$ denotes the minimum SINR requirement at the $k$-th CU. Problems (P1) and (P2) are non-convex due to the non-convex SINR constraints (i.e.,  \eqref{eq:SINR_minimum_I} and \eqref{eq:SINR_minimum_II}) and the unit-modulus constraint on reflection coefficients in \eqref{eq:phase_1}.

\section{Joint Transmit and Reflective Beamforming Design}
In this section, we propose efficient algorithms to solve the above non-convex problems based on alternating optimization, in which the transmit beamformers $\{\mathbf w_k\}$ and $\mathbf R_0$ at the BS and the reflective beamformer $\mathbf v$ at the IRS are optimized in an alternating manner. As (P1) and (P2) have similar structures, we first focus on solving (P1) in Sections III-A, III-B, and III-C, and then discuss the solution to (P2) and the differences between (P1) versus (P2) in Section III-D.

\begin{table}
\centering{
\caption{ \textbf{\upshape Algorithm 1:} Algorithm  for Solving Problem (P1)\label{tab:table1}}}
\hrule
        \begin{enumerate}[a)]
            \item Set iteration index $r= 1$ and initialize  $\mathbf{v}^{(r)}$ randomly
            \item \textbf{Repeat}: \begin{enumerate}[1)]
            				\item Under given $\mathbf{v}^{(r)}$, solve problem (SDR1.2) to obtain the optimal solution as $\{\tilde{\mathbf W}_k^{(r)}\}$ and $ \tilde{\mathbf R}_0^{(r)}$
            				\item Construct the optimal rank-one solution $\{\mathbf w_k^{\star(r)}\}$ and $ \mathbf R_0^{\star(r)}$ to (P1.1) based on $\{\tilde{\mathbf W}_k^{(r)}\}$ and $ \tilde{\mathbf R}_0^{(r)}$ by using Proposition 1
            				\item Under given $\{\mathbf w_k^{\star(r)}\}$ and $ \mathbf R_0^{\star(r)}$, solve problem (P1.4) to obtain the reflective beamformer $\mathbf{v}^{(r+1)}$
            				\item Update $r\gets r+1$
            				\end{enumerate}
            				\item \textbf{Until} convergence
        \end{enumerate}
\hrule 
\end{table}

\subsection{Transmit Beamforming Optimization at BS for (P1)}
First, we optimize the transmit beamformers $\{\mathbf w_k\}$ and $\mathbf R_0$ in (P1) under any given reflective beamformer $\mathbf v$. The transmit beamforming optimization problem is formulated as
\begin{subequations}
\begin{align} \nonumber
  \text{(P1.1)}:
       \min_{\{\mathbf w_k\}, \mathbf R_0}& \quad   \mathrm{tr}\left(\left(\mathbf G\left(\sum_{k\in\mathcal{K}} \mathbf w_k\mathbf w_k^{H} + \mathbf R_0\right)\mathbf G^{H}\right)^{-1}\right)   \\ \nonumber
  \text{s.t.} & \quad   \eqref{eq:SINR_minimum_I}, ~ \eqref{eq:power}~\text{and}~\eqref{eq:semi}.
\end{align}
\end{subequations}

We use SDR to obtain the {\it optimal} solution to problem (P1.1). Define $\mathbf  W_k = \mathbf  w_k \mathbf  w_k^{H}$, with $\mathbf  W_k \succeq \mathbf{0}$ and $\mathrm{rank}(\mathbf  W_k) \le 1$. By defining  $\mathbf  H_k = \mathbf  h_{k}\mathbf  h_{k}^{H}, \forall k \in \mathcal{K}$, problem (P1.1) is equivalently reformulated as 
\begin{subequations}
\begin{align} \nonumber 
  &\text{(P1.2)}:  \min_{\{\mathbf W_k\}, \mathbf R_0}    \mathrm{tr}\left(\left(\mathbf G\left(\sum_{k\in\mathcal{K}} \mathbf W_k + \mathbf R_0\right)\mathbf G^{H}\right)^{-1}\right)  \\   \nonumber
  &\text{s.t.}  \ \  \label{eq: SINR_W_I} \frac{1}{\Gamma_k}\mathrm{tr}(\mathbf  H_k\mathbf  W_k)-\sum_{i\in\mathcal{K},i\neq k}\mathrm{tr}(\mathbf  H_k\mathbf  W_i)-\mathrm{tr}(\mathbf  H_k\mathbf  R_0) \\
  &\qquad \ge \sigma^2_k, \forall k \in \mathcal{K}\\
  \label{eq: power_W}&  \qquad \sum_{k\in\mathcal{K}}\mathrm {tr}(\mathbf W_k) + \mathrm {tr}(\mathbf R_0)  \le P_0 \\ 
  \label{eq:W_semi}&  \qquad \mathbf R_0 \succeq \mathbf{0}, \mathbf  W_k \succeq \mathbf{0}, \forall k \in \mathcal{K}\\
  \label{eq:rank-one_W}& \qquad \mathrm{rank}(\mathbf W_k) \le 1, \forall k \in \mathcal{K}.
\end{align}
\end{subequations}

Then, we drop the rank-one constraints in \eqref{eq:rank-one_W} to get the SDR version of (P1.2), denoted by (SDR1.2). Note that problem
(SDR1.2) is a convex semi-definite program (SDP), which can be optimally solved by convex solvers such as CVX \cite{cvx}. Let $\{\tilde{\mathbf W}_k\}$ and $ \tilde{\mathbf R}_0$ denote the obtained optimal solution to (SDR1.2). Then we have the following proposition, which provides the {\it optimal} solution to (P1.1). 
\begin{proposition} \label{prop:SDR}
The SDR of (P1.2) or equivalently (P1.1) is tight, i.e., problems (P1.1), (P1.2), and (SDR1.2) have the same optimal value.  Given the optimal solution $\{\tilde{\mathbf W}_k\}$ and $ \tilde{\mathbf R}_0$ to (SDR1.2), the optimal solution to (P1.1) is
\begin{equation} \label{eq:w_new}
\mathbf w_k^\star = (\mathbf h_k^{H} \tilde{\mathbf W}_k \mathbf h_k)^{-1/2}\tilde{\mathbf W}_k \mathbf h_k,
\end{equation}
 \begin{equation} \label{eq:R_new}
\mathbf R_0^\star = \tilde{\mathbf R}_0 +  \sum_{k\in \mathcal{K}} \tilde{\mathbf W}_k- \sum_{k\in \mathcal{K}}\mathbf w_k^\star(\mathbf w_k^\star)^H.
\end{equation}
\end{proposition}
\begin{IEEEproof}
See Appendix A.
\end{IEEEproof}

\subsection{Reflective Beamforming Optimization at IRS for (P1)}
Next, we optimize the reflective beamformer $\mathbf v$ in (P1) under any given transmit beamformers $\{\mathbf w_k\}$ and $\mathbf R_0$. As $\text{CRB}(\mathbf H)$ in \eqref{eq:CRB_extended_R_x} is independent of the reflective beamformer $\mathbf v$, (P1) is reduced to following feasibility-check problem. 
\begin{subequations}
\begin{align} \nonumber
  \text{(P1.3)}:     \text{Find}&  \  \ \mathbf v  \\ \nonumber
  \text{s.t.} & \ \  \eqref{eq:SINR_minimum_I}~\text{and}~\eqref{eq:phase_1}.
\end{align}
\end{subequations}
To achieve a better converged solution and motivated by the design in \cite{8811733}, we further transform (P1.3) into the following optimization problem (P1.4) with an explicit objective. 
\begin{subequations}
\begin{align} \nonumber
  &\text{(P1.4)}:     \max_{\mathbf v}  \  \ \sum_{k\in\mathcal{K}}\beta_k  \\  \nonumber
  &\text{s.t.}  \ \ |\mathbf h_k^{H} \mathbf w_k|^2-\Gamma_k\sum_{i\in\mathcal{K},i\neq k}|\mathbf h_k^{H}\mathbf w_i|^2- \Gamma_k\mathbf h_k^{H}\mathbf R_0 \mathbf h_k\\ \nonumber
  &\qquad -\Gamma_k\sigma_k^2 \ge \beta_k, ~ \beta_k \ge 0, \forall k \in \mathcal{K}, ~\text{and}~ \eqref{eq:phase_1}.
\end{align}
\end{subequations}
Note that (P1.4) has a similar structure as the reflective beamforming design problem in \cite[(48)]{8811733}, which can be solved by using SDR together with Gaussian randomization, for which the details are omitted for brevity.

\subsection{Complete Algorithm for Solving (P1)}
By combining the transmit and reflective beamforming designs in Sections III-A and III-B, together with the alternating optimization, we have the complete algorithm to solve (P1), which is summarized as Algorithm~$1$ in Table I.
Notice that in each iteration of Algorithm~$1$, (P1.1) is optimally solved and leads to a non-increasing CRB value, and (P1.4) does not change the CRB value. As a result, the convergence of Algorithm~$1$ for solving (P1) is ensured.

\subsection{Algorithm for Solving (P2)}
In this subsection, we consider problem (P2) with Type-II CU receivers, which can be solved similarly as Algorithm 1 for (P1) based on  alternating optimization. Therefore, we only need to focus on the transmit and reflective beamforming in the following.

First, consider the transmit beamforming optimization problem, which is given by 
\begin{subequations}
\begin{align} \nonumber
  \text{(P2.1)}:
       \min_{\{\mathbf w_k\}, \mathbf R_0}& \quad   \mathrm{tr}\left(\left(\mathbf G\left(\sum_{k\in\mathcal{K}} \mathbf w_k\mathbf w_k^{H} + \mathbf R_0\right)\mathbf G^{H}\right)^{-1}\right)   \\ \nonumber
  \text{s.t.} & \quad    \eqref{eq:power}, ~\eqref{eq:semi},~\text{and}~ \eqref{eq:SINR_minimum_II}.
\end{align}
\end{subequations}
Problem (P2.1) can be optimally solved by using SDR. Similarly as (SDR1.2), We express the SDR of (P2.1) (after dropping the rank-one constraints) as 
\begin{subequations}
\begin{align} \nonumber 
  &\text{(SDR2.1)}:  \min_{\{\mathbf W_k\}, \mathbf R_0}    \mathrm{tr}\left(\left(\mathbf G\left(\sum_{k\in\mathcal{K}} \mathbf W_k+ \mathbf R_0\right)\mathbf G^{H}\right)^{-1}\right)  \\  
  &\text{s.t.}  \ \  \label{eq: SINR_W_II} \frac{1}{\Gamma_k}\mathrm{tr}(\mathbf  H_k\mathbf  W_k)\!-\hspace{-0.5em} \sum_{i\in\mathcal{K},i\neq k}\mathrm{tr}(\mathbf  H_k\mathbf  W_i)\ge \sigma^2_k, \forall k \in \mathcal{K} \tag{12}\\\nonumber
  & \qquad \eqref{eq: power_W}~\text{and}~\eqref{eq:W_semi}.
\end{align}
\end{subequations}
Let $\{\bar{\mathbf W}_k\}$ and $\bar{\mathbf R}_0$ denote the optimal solution to (SDR2.1). Then we have the following proposition.
\begin{proposition} \label{prop:SDR_2}
Problems (P2.1) and (SDR2.1) have the same optimal value. The  optimal solution $\{\mathbf w_k^{\star\star}\}$ and $\mathbf R_0^{\star \star}$ to problem (P2.1) can be constructed based on $\{\bar{\mathbf W}_k\}$ and $\bar{\mathbf R}_0$ to (SDR2.1) similarly as in \eqref{eq:w_new} and \eqref{eq:R_new}.
\end{proposition}
\begin{IEEEproof}
See Appendix B.
\end{IEEEproof}

Next, consider the reflective beamforming optimization problem, which is same as (P1.3) by setting $\mathbf R_0 =\mathbf 0$,  denoted by (P2.2). This problem can be solved based on the design in Section III-B, for which the details are omitted. Therefore, by combining the solutions to (P2.1) and (P2.2) together with alternating optimization, problem (P2) is finally solved. 

\begin{remark}
By comparing problems (P1) and (P2), it is clear that the optimal value of (P2) serves as a lower bound on that of (P1), because every feasible solution to (P1) is also feasible to (P2) but not vice versa. This shows the benefit of sensing signal interference pre-cancellation at Type-II CU receiver. Nevertheless, it is worth pointing out that for the special case with one CU, Type-II CU receivers do not provide performance gains over its Type-I counterpart, as shown in the following proposition.
\end{remark}
\begin{proposition} \label{prop:single-user}
When $K=1$, the optimal values achieved by  problems (P1) and (P2) are identical.
\end{proposition} 
\begin{IEEEproof}
See Appendix \ref{sec:proof_of_proposition_2}.
\end{IEEEproof}

\section{Numerical Results}
In this section, we provide numerical results to validate the performance of our proposed joint beamforming design. We consider the distance-dependent path loss model, i.e., $L(d)=K_0(d/d_0)^{-\alpha_0}$, where $K_0=-30$~dB is the pathloss at the reference distance $d_0 =1$~m and $\alpha_0$ is the path loss exponent. We set $\alpha_0$ as $2.2$, $2.2$, and $3.5$ for the BS-IRS, IRS-CU, and BS-CU links, respectively. We consider the Rician fading for the BS-IRS and BS-CU links with the Rician factor being $0.5$, and Rayleign fading for the IRS-CU link. Also, additional shadow fading is considered for the BS-CU links, with a standard deviation of $10~ \text{dB}$. The BS, the IRS, and the three CUs are located at coordinate $(0,0)$, $(4~\text{m},2~\text{m})$, $(50~\text{m},0~\text{m})$, $(45~\text{m},-2~\text{m})$, and $(55~\text{m},-2~\text{m})$, respectively. Without loss of generality, we assume that all users have the same SINR requirements, i.e., $\Gamma_k = \Gamma, \forall k \in \mathcal{K}$. We set $M =8$, $N =8$, $T=256$, $P_0 =30~\text{dBm}$, $\sigma_\text{R}^2 = -110~\text{dBm}$, and $\sigma_k^2 = -80~\text{dBm}, \forall k \in \mathcal{K}$.

For performance comparison, we consider the following benchmark schemes.

\subsubsection{Transmit beamforming only with random phase shifts (Transmit BF only)} This scheme optimizes the transmit beamformers $\{\mathbf w_k\}$ and $\mathbf R_0$ at the BS under given random phase shifts at IRS. This method is implemented by using (SDR1.2) and Proposition 1 (or (SDR2.1) and Proposition 2) for the case with Type-I (II) CU receivers.

\subsubsection{Separate information and sensing beamforming design (Separate BF design)} This scheme optimizes the transmit information and sensing beamformers separately. First, we optimize the transmit information beamformers $\{\bar{\mathbf w}_k\}$ and the reflective beamformer $\mathbf v$ to minimize the transmit power while ensuring the SINR constraints at the CUs\cite{8811733}, and accordingly set $\mathbf w_k = \alpha \bar{\mathbf w}_k, \forall k\in \mathcal{K}$, with $\alpha\ge 1$.  Then, we optimize the transmit sensing beamformer $\mathbf R_0$ together with $\alpha$ to minimize the estimation CRB, subject to the minimum SINR constraints at the CUs and the maximum transmit power constraint at the BS.

\begin{figure}[t]
    \centering
    \includegraphics[width=0.38\textwidth]{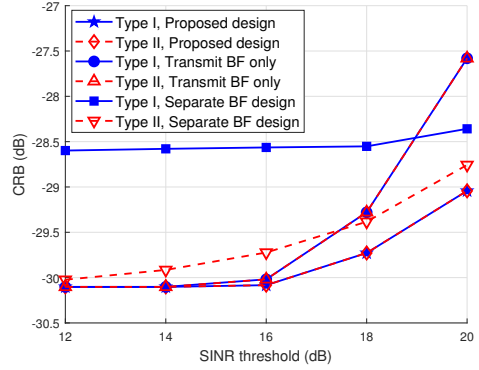}
    \caption{The estimation CRB versus the SINR threshold $\Gamma$ with $K=1$.}
    \label{fig:CRB_SNR_single}
\end{figure}
\begin{figure}[t]
    \centering
    \includegraphics[width=0.38\textwidth]{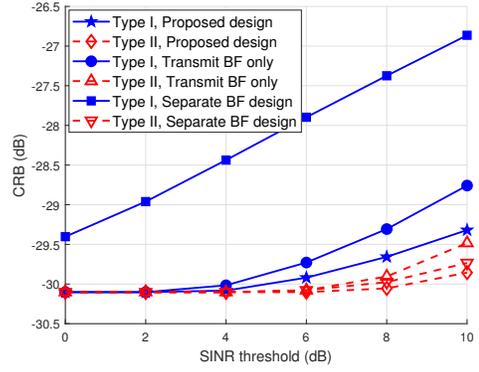}
    \caption{The estimation CRB versus the SINR threshold $\Gamma$ with $K=3$.}
    \label{fig:CRB_SNR_multi}
\end{figure}

Fig.~\ref{fig:CRB_SNR_single} shows the estimation CRB versus the SINR threshold $\Gamma$ with $K=1$. It is observed that for both types of CU receivers, the proposed designs achieve the same CRB performance, and outperform other benchmark schemes. This is consistent with Proposition~\ref{prop:single-user} and shows the benefit of our proposed design. When the SINR threshold is low, the transmit BF only is observed to outperform the separate BF design. This is due to the fact that the estimation CRB only depends on the transmit beamforming, which is thus crucial in this case.  When the SINR threshold is high, the separate BF design is observed to outperform the transmit BF only. This is because that the communication performance is the performance bottleneck in this case, which critically relies on the reflective beamforming. 

Fig.~\ref{fig:CRB_SNR_multi} shows the estimation CRB versus the SINR threshold $\Gamma$ with $K=3$. It is observed that the proposed design achieves the lowest CRB for each type of CU receivers, and the case with Type-II CU receivers leads to much lower CRB than that with Type-I CU receivers. This validates the benefit of sensing signal interference pre-cancellation. It is also observed that the transmit BF only outperforms the separate BF design for Type-I CU receivers, but the opposite is true for Type-II CU receivers. This shows that the joint information and sensing transmit beamforming design becomes more important when the interference from sensing signals becomes harmful for communication.

\section{Conclusion}
This paper studied the joint transmit and reflective beamforming design in an IRS-enabled multiuser ISAC system with one extended target. We considered that the BS sent dedicated sensing signals in addition to information signals, and accordingly considered two types of CU receivers without and with the sensing signal interference pre-cancellation capability, respectively. We proposed to minimize the estimation CRB by jointly optimizing the transmit and reflective beamforming, subject to the minimum SINR constraints at the CUs and maximum transmit power constraint at the BS. Numerical results showed that the joint beamforming design achieved enhanced ISAC performance in terms of lower CRB than other benchmarks without such joint consideration, and the dedicated sensing signals together with interference pre-cancellation was crucial for further performance improvement. 

\appendix
\subsection{Proof of Proposition \ref{prop:SDR}} \label{sec:proof_of_proposition_1}
Define $\mathbf  W_k^\star = \mathbf  w_k^\star ( \mathbf w_k^\star)^{H} \succeq \mathbf 0, \forall k \in \mathcal{K}$. We prove this proposition by showing that $\{\mathbf W_k^\star\}$ (of rank one) and $\mathbf R_0^\star$ are also optimal for (SDR1.2). It follows from \eqref{eq:R_new} that $ \sum_{k\in\mathcal{K}} \mathbf W_k^\star + {\mathbf R}_0^\star =  \sum_{k\in\mathcal{K}}\tilde{\mathbf W}_k+\tilde{\mathbf R}_0$. As a result, the objective value and transmit power obtained by $\{\mathbf W_k^\star\}$ and $ \mathbf R_0^\star$ remains the same. It can be verified that $\mathrm {tr}(\mathbf H_k  \mathbf W_k^\star)=\mathrm {tr}(\mathbf H_k \tilde{\mathbf W}_k), \forall k \in \mathcal{K}$.
Therefore,  the SINR constraint in \eqref{eq: SINR_W_I} is met.
Furthermore, for any $\mathbf y \in \mathbb{C}^{M \times 1}$, it holds that
\begin{equation}\nonumber
\mathbf y^{H} (\tilde{\mathbf W}_k - \mathbf W_k^\star)\mathbf y = \mathbf y^{H} \tilde{\mathbf W}_k \mathbf y - |\mathbf y^H \tilde{\mathbf W}_k \mathbf h_k|^2(\mathbf h_k^H \tilde{\mathbf W}_k \mathbf h_k)^{-1}.
\end{equation}
According to the Cauchy-Schwarz inequality, we have
$(\mathbf y^H \tilde{\mathbf W}_k \mathbf y) (\mathbf h_k^{H} \tilde{\mathbf W}_k \mathbf h_k)   \geq |\mathbf y^H \tilde{\mathbf W}_k \mathbf h_k|^2$,
and thus it follows that $\mathbf y_k^H (\tilde{\mathbf W}_k - \mathbf W_k^\star)\mathbf y\geq 0$. Accordingly, we have  $\tilde{\mathbf W}_k - \mathbf W_k^\star \succeq \mathbf 0,\forall k \in \mathcal{K}$. It follows from \eqref{eq:R_new} that $\mathbf R_0^\star \succeq \mathbf 0$. Hence, $\{\mathbf W_k^\star\}$ and $ \mathbf R_0^\star$ are optimal for (SDR1.2).  Proposition \ref{prop:SDR} is proved.

\subsection{Proof of Proposition \ref{prop:SDR_2}} \label{sec:proof_of_proposition_1_2}
Define $\mathbf  W_k^{\star\star} = \mathbf  w_k^{\star\star} ( \mathbf w_k^{\star\star})^{H}\succeq \mathbf 0, \forall k \in \mathcal{K}$. We prove this proposition by showing that $\mathbf  W_k^{\star\star}$ (of rank one) and $\mathbf R_0^{\star \star}$ are also optimal for (SDR2.1). 
Similar to (SDR1.2), the objective value remains the same for the reconstructed solution and the constraints in \eqref{eq: power_W} and~\eqref{eq:W_semi} are met. We have $\mathrm {tr}(\mathbf H_k \mathbf W_k^{\star\star} )
=\mathrm {tr}(\mathbf H_k \bar{\mathbf W}_k)$ and $\bar{\mathbf W}_k- \mathbf W_k^{\star\star} \succeq \mathbf 0, k \in \mathcal{K}$,
 such that
 \setcounter{equation}{12} 
\begin{equation}
\begin{split}
&\frac{1}{\Gamma_k}\mathrm{tr}(\mathbf  H_k\mathbf W_k^{\star\star} )-\sum_{i\in\mathcal{K},i\neq k}\mathrm{tr}(\mathbf  H_k\mathbf W_i^{\star\star} )\\
\ge&\frac{1}{\Gamma_k}\mathrm{tr}(\mathbf  H_k\bar{\mathbf W}_k)-\sum_{i\in\mathcal{K},i\neq k}\mathrm{tr}(\mathbf  H_k\bar{\mathbf W}_i)
\ge\sigma_k^2.
\end{split}
\end{equation}
Then $\mathbf  W_k^{\star\star}$ and $\mathbf R_0^{\star \star}$ also satisfy the SINR constraint in \eqref{eq: SINR_W_II}, and thus are optimal for (SDR2.1).  Proposition \ref{prop:SDR_2} is proved.

\subsection{Proof of Proposition \ref{prop:single-user}} \label{sec:proof_of_proposition_2}
When $K=1$, it is easy to show that (SDR2.1) has an optimal solution satisfying $\bar{\mathbf R}_0 =\mathbf 0$, as the SINR constraint \eqref{eq: SINR_W_II} is independent of $\bar{\mathbf R}_0$ and the objective value is only related to the sum of $\bar{\mathbf R}_0 $ and $\bar{\mathbf W}_1$. Then, the optimal solution to (SDR2.1), i.e.,  $\bar{\mathbf R}_0=\mathbf 0$ and $ \bar{\mathbf W}_1$, is also optimal to (SDR1.2). Then,  based on Propositions \ref{prop:SDR} and \ref{prop:SDR_2},  (P1.1) and (P2.1) have the same optimal objective value under any same reflective beamformer $\mathbf v$. As a result, the optimal values of (P1) and (P2) are identical. Proposition \ref{prop:single-user} is proved.

\ifCLASSOPTIONcaptionsoff
  \newpage
\fi

\bibliographystyle{IEEEtran}
\bibliography{IEEEabrv,mybibfile}

\end{document}